# On the geometry optimization of X-band resonator for electron paramagnetic resonance application

Mikhail Y. Ivanov

*International Tomography Center, SB RAS, Novosibirsk 630090, Russia*

michael.ivanov@tomo.nsc.ru

The microwave resonator is one of the key components in the modern Electron Paramagnetic Resonance (EPR) spectroscopy setup, as it largely determines the performance characteristics and limitations of the entire spectrometer. In this research note the possible way of resonator optimization is described. A detailed computer model describing the distribution of the electromagnetic field for Bruker 4118X-MD-5W1 design resonator has been developed. All details including a dielectric insert, conductive screen, coupling antenna and PTFE supports were included in the model. All dissipation processes were considered in the calculation. The influence of the resonator geometry on the Q-factor, filling factor and operating frequency of the resonant mode has been investigated. The resonator geometric parameters are optimized to achieve maximum sensitivity. According to the calculations, the optimized resonator structure has in 10 times greater sensitivity than the original MD-5 design.

**Introduction**

The electron paramagnetic resonance (EPR) spectroscopy provides a unique information about the paramagnetic centers – centers containing unpaired electrons. As in many other methods, the range of actual research and applied problems solved by the EPR is limited by the sensitivity of the spectrometer. The requirement for high sensitivity is particularly acute when working with biological samples, where the accumulation of an acceptable signal/noise ratio may require several weeks of continuous operation of the experimental setup. One of the most effective methods of improving the EPR setup sensitivity is the modernization of the microwave resonator.

The use of resonators in experimental technology significantly expands the functionality of traditional spectroscopic techniques. In the field of EPR spectroscopy, resonators allow focusing microwave energy on the sample and spatially separate the electrical and magnetic components of the electromagnetic field that significantly increases the sensitivity compared to non-resonator EPR techniques. In dielectric resonators the dielectric insert is located inside the conductive screen. Together, the dielectric and the screen form a resonant system.

The growing interest in the study of biological systems sets increasingly high requirements for the sensitivity of the spectrometer, which leads to the need to further increase the sensitivity of resonators through the use of new technologies and microwave design. A significant amount of research is devoted to the problem of increasing the sensitivity of the spectrometer by modifying the design of the dielectric resonator: dielectric resonators for high-pressure EPR setup [1,2], multi-sample EPR cavity [3], double-stacked and ceramic resonators [4,5]. In addition, there are also some theoretical studies devoted to the EPR resonator optimization [6–9].

Among the variety of EPR dielectric resonators, the Bruker ER 4118X-MD-5W1 (MD-5) is one of the most widely distributed and broadly used. This resonator is based on the alumina dielectric insert and has an advantage of high Q-value (>20000, [10]) and large sample access (5 mm). However, if the amount of available sample is limited, what is common in biological applications, the big volume of MD-5 resonator leads to a low filling factor and low sensitivity. In this research note we described the possible way of optimization the current MD-5 design to achieve better sensitivity for EPR experiments with the small amount of samples.

**Resonance frequency**

Figure 1a shows a general view of the model with the indication of resonator components and the mesh used to solve Maxwell's equations by the finite element method implemented in the Comsol Multiphysics® package. To improve the calculation quality in critical areas, the grid was additionally densified in the area of the dielectric insert and the location of the sample (this area is highlighted by blue in figure 1a). To reduce the calculation time, the solution was carried out for the half of real resonator model taking into account the mirror symmetry of the system. For the simplicity, figure 1a does not show the auxiliary sphere in the region of the cavity hole, which is used to calculate the radiation losses, also some elements of the conductive screen and PTFE inserts are hidden. The model takes into account all sources of energy losses: surface currents in the conductive parts - the screen and coupling antenna, dielectric losses and radiation losses through the end holes in the screen. Figure 1b shows the four geometric parameters that were varied to achieve maximum sensitivity. There are inner ($D_{in}$) and outer ($D_{out}$) diameters of dielectric insert, the height of insert (H) and the diameter of conductive screen ($D_{screen}$). During the computational research only one parameter had been varied and others were fixed. By default Bruker MD-5 resonator has the following dimensions: $D_{out}$ = 10 mm, $D_{in}$ = 5 mm, H = 13 mm, $D_{screen}$ = 16.4 mm.

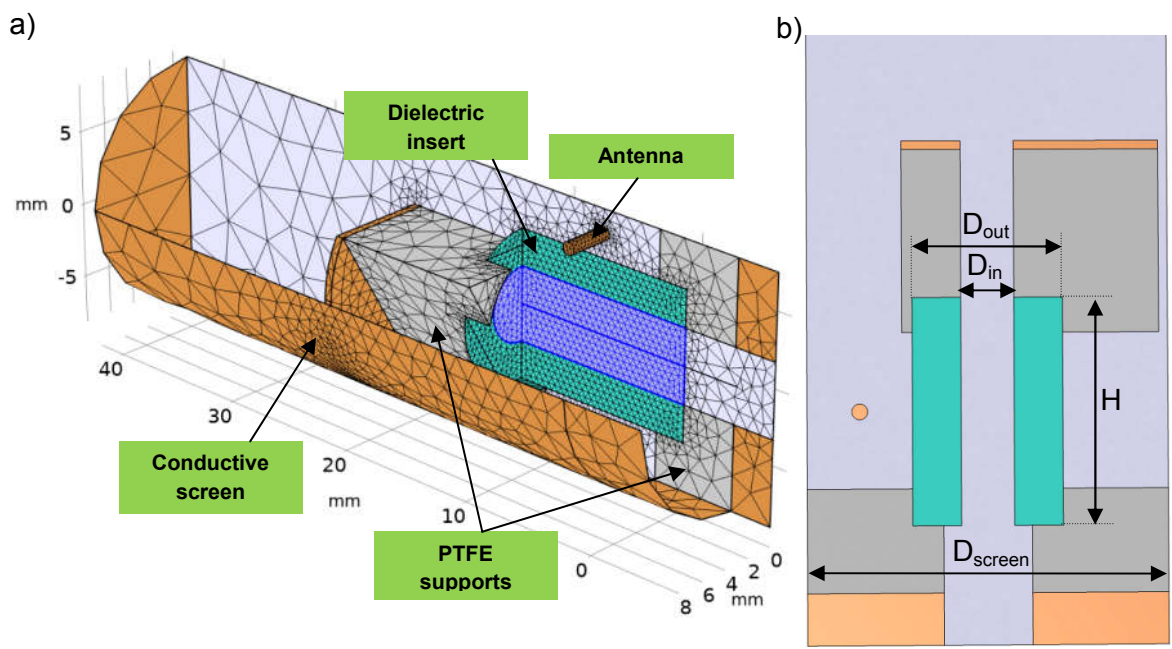

Fig. 1. a) General model view with generated mesh. The volume filled with the sample is highlighted by blue. b) Model side view. The varied dimensions of dielectric insert and screen are indicated.

Figure 2 shows the distribution of the electromagnetic field components for the main operating mode: $TE_{01\delta}$. Of the features of this mode, it is worth noting the low concentration of the electrical component inside the dielectric insert and the high uniformity of the magnetic component along the inner volume of dielectric insert [11,12]. These features are extremely important for EPR applications. The EPR sample is placed in the central hole of the dielectric insert and the low concentration of the electrical component on the sample makes possible to provide experiments with polar solvents (i.e. water-glycerol and water-ethanol mixtures). A uniform magnetic field is especially critical for pulse EPR measurements where the overall sample magnetization is being rotated by microwave pulses [13–15].

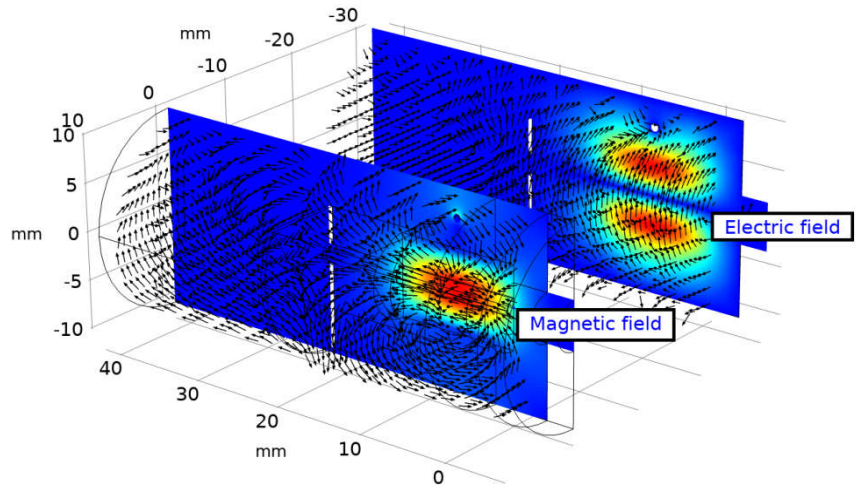

Fig. 2. The distribution of magnetic and electric components for $TE_{01\delta}$ resonance mode. Color indicates the relative intensity of the corresponding field. Arrows depict the direction of the field.

Firstly, we were interested in the resonant frequency dependence on the dielectric insert dimensions. Parameters $D_{out}$, $D_{in}$ and $H$ were indepentendly varied in the corresponding dimension range. The position of $TE_{01\delta}$ resonant frequency was calculated. Figure 3 collects the results of this computational research.

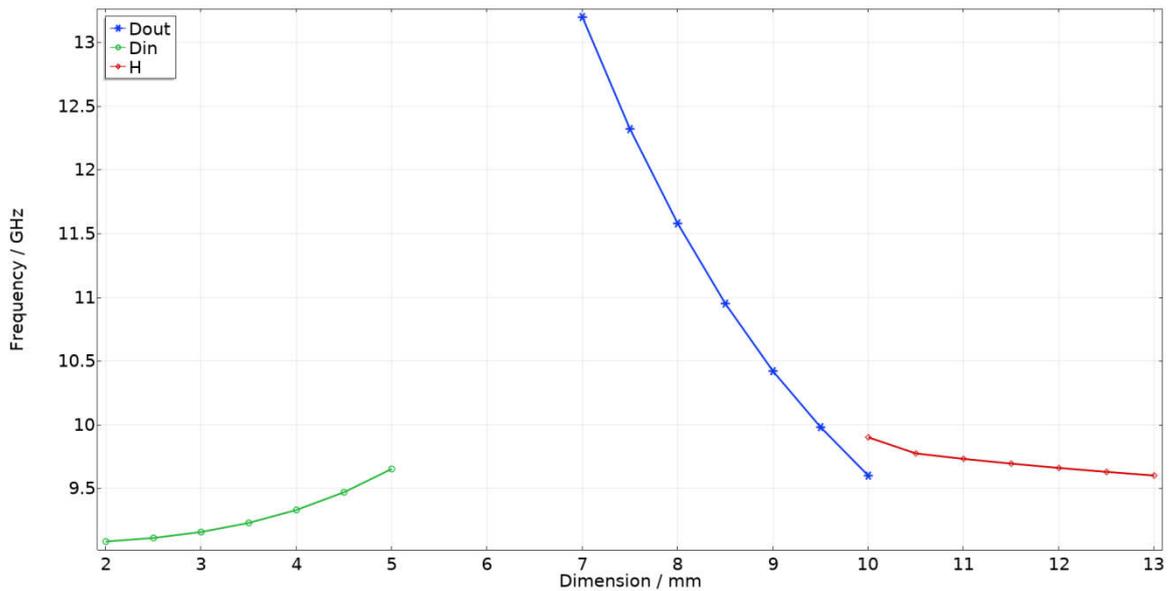

Fig. 3. The $TE_{01\delta}$ resonant frequency dependence on the dielectric insert dimensions $D_{out}$ (blue line), $D_{in}$ (green line) and $H$ (red line). Color lines are guides for the eye.

For the $D_{out}$ - curve a monotonic frequency drop with a characteristic frequency response 1.2 GHz/mm is observed (the blue curve at figure 3), the relative frequency change is ~37%. The $D_{in}$ - curve (green curve at figure 3), in contrast, demonstrates a monotonic increase in the resonance frequency. It is worth noting that the frequency response here is 0.16 GHz+/mm, what is an order of magnitude less than the variation of the outer diameter, the relative frequency change here is about 5 %. For the H - curve (red curve at figure 3) there is a monotonic frequency drop with a characteristic step of 0.1 GHz/mm and 3% frequency change. Thus, the position of the resonant frequency is an order of magnitude more sensitive to the $D_{out}$ variation than to other parameters of dielectric insert. This statement is not obvious from the general considerations for this geometry.

Figure 4 shows the dependence of the Q-factor for operating mode on the varying the geometric dimensions of the dielectric insert. There is a monotonic Q-factor drop while $D_{in}$ increases (green curve at figure 4), the difference in absolute value is ~10% with a diameter change from 2 mm to 5 mm. Varying the insert height (red curve at figure 4), the Q-factor does not significantly change that can be explained by the axial symmetry of the conductive currents in the shield and their weak dependence on the longitudinal size of the resonator (see figure 2). Red line reaches a constant value Q~25000 at the point H = 11 mm.

While increasing $D_{out}$ (blue curve at figure 4) the Q-factor increases on ~ 15% in the absolute value. However, after the point $D_{out}$ = 8.0 mm Q-factor drops down. Such behavior can be explained by a significant losses increase caused by surface currents on the screen and antenna surface. While the outer surface of dielectric ring approaches the screen surface (the screen diameter was fixed, $D_{screen}$ = 16.4 mm), the electric field, that propagates out of the dielectric insert, induces intensive currents on the screen, that leads to a Q-factor decrease.

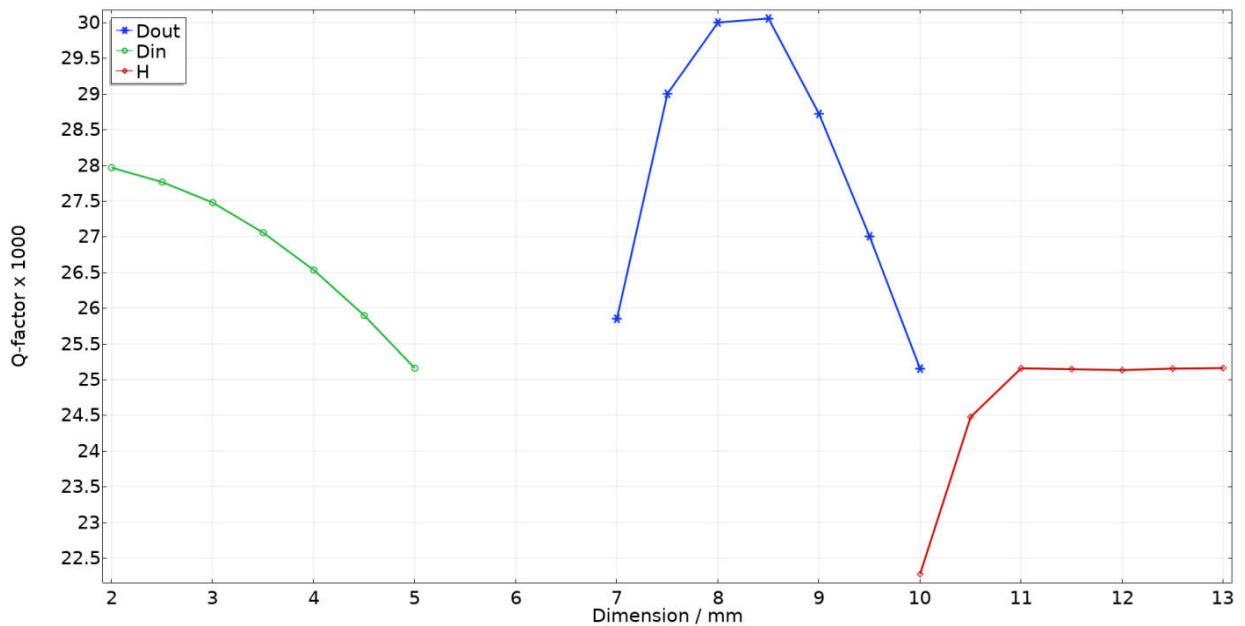

Fig. 4. The Q-factor dependence of $TE_{01\delta}$ mode on the dielectric insert dimensions $D_{out}$ (blue line), $D_{in}$ (green line) and H (red line). Color lines are guides for the eye.

This assumption is confirmed by calculations presented at figure 5. Results demonstrate that the relative fraction of conductive losses changes from ~ 76% at $D_{out}$ = 7 mm to ~ 90% at $D_{out}$ = 10 mm, while the fraction of dielectric losses falls from 24% to 11%, respectively. In this case, the radiation losses contribution is less than 1% and could be neglected. The point of the highest Q-factor at figure 4 corresponds to the optimal ratio of conductive and dielectric losses, what is achieved at the values of the $D_{out}$ ~ 8.0 mm.

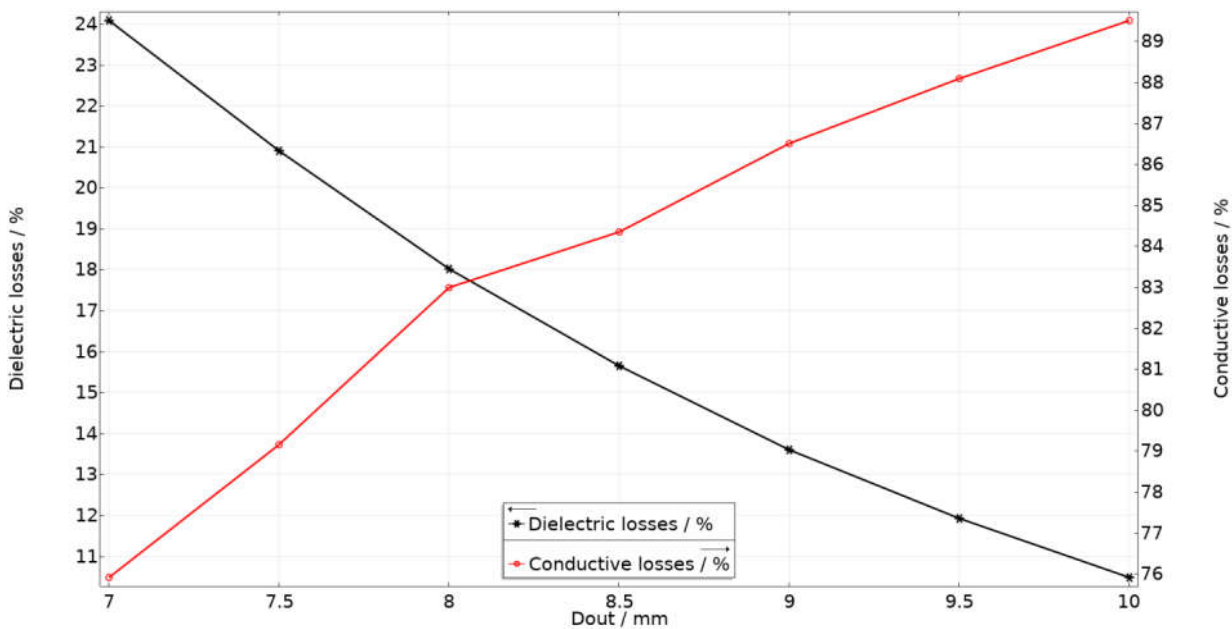

Fig. 5. The relative fraction dependence of dielectric (black line, left axes) and conductive (red line, right axes) losses on $D_{out}$. Color lines are guides for the eye.

From figure 4 one can understand the strategy of resonator design optimization. Variation of insert height has approximately no effect on the Q-factor. The optimal $D_{out}$ leads in the region 8-9 mm. The $D_{in}$ should be as less as possible, but considering the actual size of the sample quartz tubes, the minimum $D_{in}$ was chosen as 2 mm. It is important to note that resonance frequency should lie in the range of 9.4-9.8 GHz - that is the optimal range of mw source generation in commercial Bruker equipment. According to the calculations results the dielectric insert with the geometry 9/2/13 mm (here and after the designation $D_{out}$ / $D_{in}$ / H is used) has a resonance at 9.7 GHz and Q ~ 33000, what is approximately at 25% higher than the Q-value of the commercial dielectric insert with a non-optimized size 10/5/13 mm (Q ~ 25000).

The next step is the conductive screen geometry optimization. Figure 6 shows the resulting dependence of Q-factor and resonance frequency on $D_{screen}$. While the $D_{screen}$ increases, the Q-factor is also monotonically growing. Qualitatively, the increase in Q-factor is associated with a decrease in conductive losses, which in this range fall by more than 10%: from 91% to 80 %. Thus, for a maximum D$_{screen}$ = 20.4 mm, the Q-factor reaches 37000. In practice, maximum screen size is limited by the distance between electromagnet poles and the inner diameter of cryostat where the resonator stands.

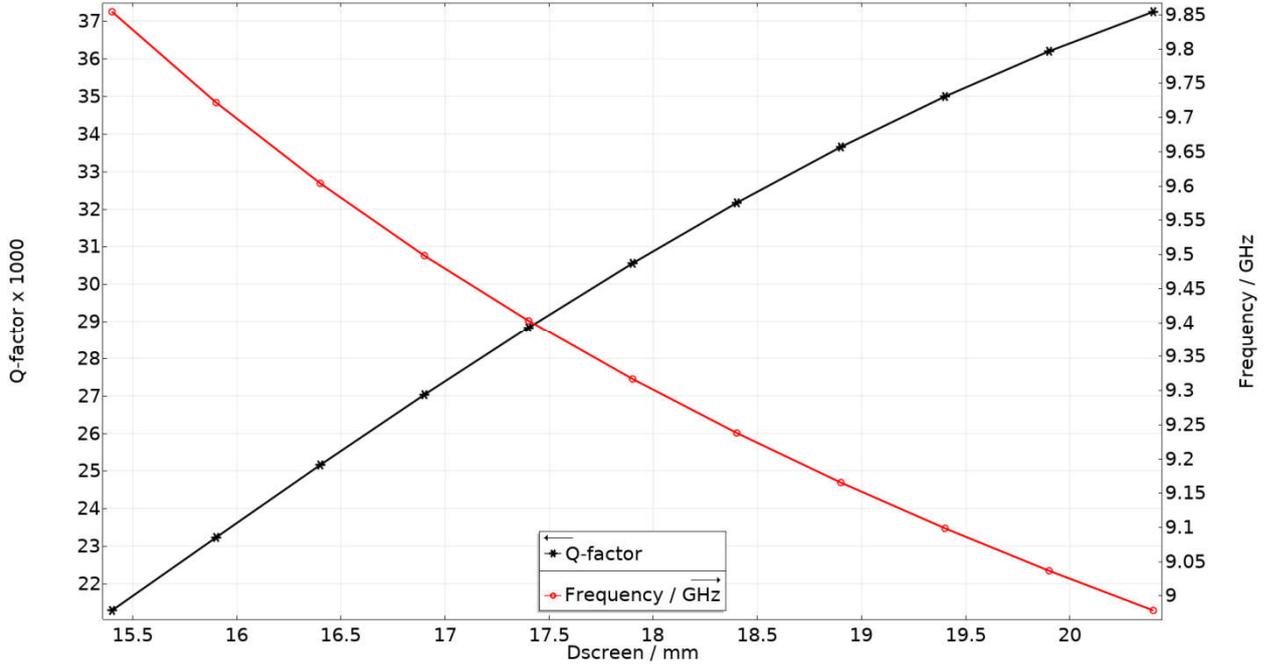

Fig. 6. The Q-factor (black line, left axes) and resonance frequency (red line, right axes) dependence on the D$_{screen}$. Color lines are guides for the eye.

To combine the optimization results for dielectric insert dimensions and screen geometry we concluded that the insert with 8.5/2/13 sizes and D$_{screen}$ = 20.4 mm provide resonance at 9.58 GHz and has Q ~ 44000, what is in 80% higher compared to the original MD-5 geometry without optimization.

**Filling factor**

The filling factor is an another important parameter of resonator that determines EPR-setup sensitivity. The filling factor is critical in problems with a limited sample size, as is in the case of biological EPR applications.

The filling factor is defined as

$$\eta = 100\% * \frac{\int_{V_{sample}} |B^2(\vec{r})| dV}{\int_{V_{all}} |B^2(\vec{r})| dV}$$

It is the ratio of stored microwave energy in the sample volume relative to the energy stored in the entire resonator. To carry out numerical calculations sample tube with 1.8 mm outer diameter was implemented. The $\eta$ dependence on the dielectric insert geometry is shown at figure 7. The change in the $D_{in}$ has the greatest influence on the filling factor, $\eta$ increases from 2.6% to 4.5%. Also, the filling factor increases with a decrease in the $D_{out}$ - up to 3.8 %. The H variation has almost no effect on $\eta$.

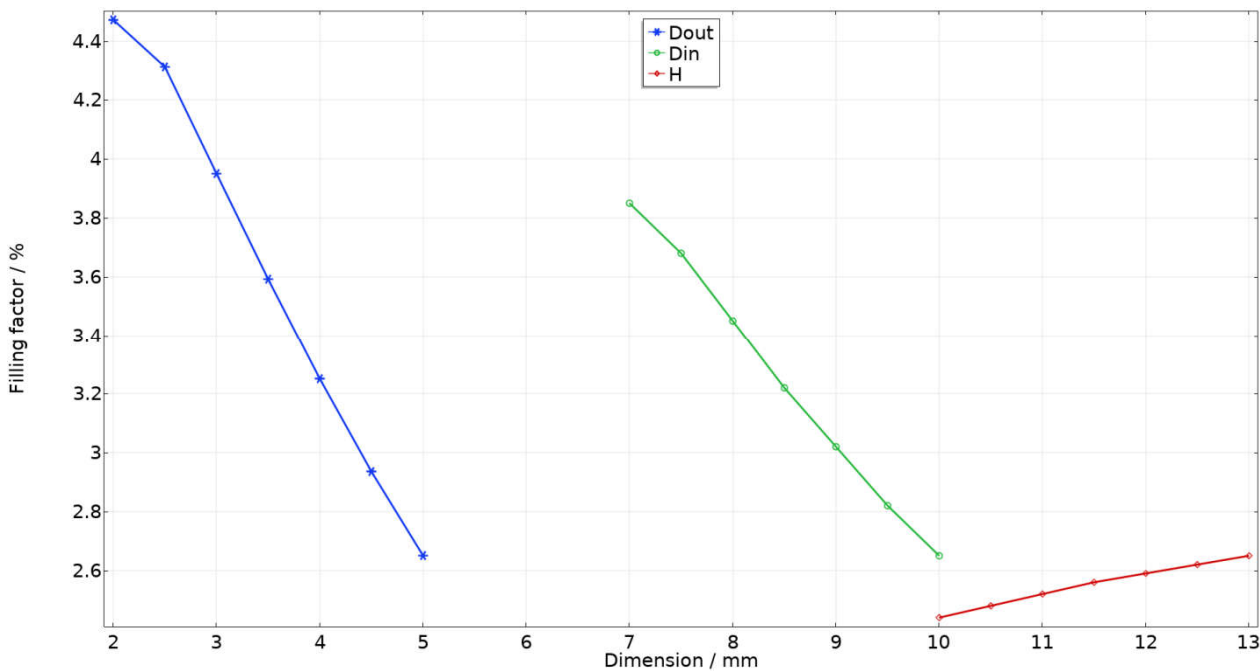

Fig. 7. The filling factor dependence on the dielectric insert dimensions $D_{out}$ (blue line), $D_{in}$ (green line) and H (red line). Color lines are guides for the eye.

As shown above, in terms of Q-factor, the optimal insert size is 8.5/2/13. However, according to figure 7, this size is also optimal in terms of the filling factor.

Thus, compared to the non-optimized dielectric insert, which has a filling factor of 2.6 %, the optimized insert has a filling factor of 5.4 %. Since the final amplitude of EPR signal depends on the Q-factor and filling factor (for a fixed sample amount), then, in accordance with the calculations, the optimized resonator with a screen diameter of 20.4 mm and dielectric insert size 8.5/2/13 has Q ~ 44000 and $\eta$ ~ 5.4 %. That should give a 3.5 signal gain. In the case of white noise final signal-noise ratio is proportional to the squared accumulation time. Thus, the modification of the resonator using the optimized geometry of dielectric insert and the screen will increase the sensitivity of the spectrometer (or decrease of the accumulation time for a fixed signal/noise ratio) by an order of magnitude.

**Acknowledgements**

This work was supported by the RFBR (No. 18-38-00394).